\documentstyle[prb,aps,preprint]{revtex}
\draft
\input{psfig}
\begin{document}
\title{Protein Design in a Lattice Model of Hydrophobic and Polar Amino
Acids}
\author{Cristian Micheletti$^1$, Flavio Seno$^1$, Amos Maritan$^2$ 
and Jayanth R. Banavar$^3$}
\vskip 0.3cm
\address{(1)INFM-Dipartimento di Fisica, Universit\`a di Padova, 
Via Marzolo 8,35131  Padova, Italy }
\address{(2) International School for Advanced Studies (S.I.S.S.A.),
Via Beirut 2-4, 34014 Trieste, Italy }
\address{(3) Department of Physics and Center for Materials Physics,
104 Davey Laboratory, The Pennsylvania State University, University
Park, Pennsylvania 16802}
\date{\today}
\maketitle

\begin{abstract}
A general strategy is described for finding which amino acid
sequences have native states in a desired conformation (inverse
design).  The approach is used to design sequences of 48 hydrophobic
and polar aminoacids on three-dimensional lattice structures.
Previous studies  employing a sequence-space Monte-Carlo
technique  resulted in the successful design of one sequence
in ten attempts.  The present work also entails the exploration of
conformations that compete significantly with the target structure
for being its ground state. The design procedure is successful in all 
the ten cases.
\end{abstract}


 A formidable challenge in molecular biology is the suc\-ces\-sful
design \cite{1,2,3,4,5,6,7,8,9,10,11,12,13,14,15} of sequences of
amino acids that fold rapidly into desired native conformations
commonly assumed to be their ground states -- protein functionality is
controlled by the native state structure.  It has been recognized that
a simple binary pattern of hydrophobic and hydrophilic residues along
the polypeptide chain encode structure at the coarse-grained level
\cite{1,5}. Thus the simplest model of proteins consists of sequences
made up of just two kinds of amino acids (H and P representing
hydrophobic and polar residues) configured as self-avoiding chains on
a lattice and described by a contact Hamiltonian \cite{16,17,18}. Such
models are known to adequately describe proteins at the coarse-grained
level with the advantage that the native states can be determined
exactly \cite{16,17,18,19,20,21,22,23}. Furthermore, they provide a
controlled laboratory for theoretical investigations and rigorous
testing of concepts and ideas for future use in studies on real
proteins. Within this framework, a Harvard-San Francisco team [HSF]
\cite{6} recently carried out tests of the design of three-dimensional
cubic lattice heteropolymers of length 48. Ten maximally compact
conformations (see table I) were chosen as target structures and
attempts were made to design sequences that would have these as the
ground state.  Disappointingly, nine out of the ten designed sequences
were found to have ground states in conformations other than the
target structures.  The HSF study is the most stringent test, to date,
of protein design procedures -- it considers the longest designed
sequences in three dimensions whose true ground state could yet be
determined rigorously.

In this letter we present and discuss a novel inverse design approach 
for three-dimensional HP lattice proteins. The method, which encompasses
negative design features, is found to be both reliable and efficient in
isolating sequences which fold into a given target conformation.
For each of the 10 HSF cases, we have confirmed, using the constraint-based 
hydrophobic core construction (CHCC) method of Yue and Dill \cite{24,25}, 
that our design strategy is successful.

According to the standard HP model \cite{16,17,18} the energy of a
sequence in a lattice conformation is simply given by the negative of
the number of contacts between pairs of H residues which are not
consecutive in the chain.  The contact Hamiltonian can be simply
written as

\begin{equation}
{\cal H} = -{1 \over 2} {\tilde{\sum}}_{\langle i,j \rangle} S_i S_j
\delta(r_{ij},1) 
\label{eqn:ham}
\end{equation}

\noindent where $S_i=0$ $[S_i=1]$ denotes the polarity
[hydrophobicity] of residue $i$, $r_{ij}$ is the distance between
residues $i$ and $j$ measured in lattice spacings and the tilde is
used to indicated that the sum does not include pairs of residues
which are  consecutive along the chain.  The native state of a
sequence ${\cal S} = \{S_1, S_2, ...  \}$ is the ground state
conformation of the contact Hamiltonian.  For the HSF problem, the
inverse design entails the identification of the sequence (or one of
the sequences) among the $2^{48}$ ($ > 10^{14}$) sequences that has a
native state in the target structure (which is one of approximately
$10^{32}$
self-avoiding conformations for a chain of length 48 on a cubic
lattice). Typically, because there are just two kinds of amino acids in
the HP lattice model,
there exist several sequences that solve the design problem for a
given structure. In general, these solutions will not be equivalent in
terms of thermodynamic stability. For example, sequences containing at
most one H residue admit all possible structures as native states (the
ground state energy being always zero); hence they represent trivial
solutions to all design problems. This occurs at expenses of
thermodynamic stability: due to the tremendous degeneracy of their
native states, the probability that these sequences are found in a
given target conformation is vanishingly small even at zero
temperature.
Another example of a trivial solution is the sequence
consisting of no $P$ residues which, indeed, admits all compact
conformations as native states.

The goal of a design procedure is to isolate the solutions with lowest
possible degeneracy which ensures the highest low-temperature
occupation of the target conformation.  Stated
mathematically\cite{10,11,12}, in order to perform an inverse design
on a target structure, $\Gamma$, one needs to identify the one (or many),
$S$, that maximizes the occupation probability according to Boltzmann
statistics,

\begin{equation}
P_\Gamma(S) = { e^{ - \beta E_S(\Gamma)} \over \sum_{\Gamma^\prime}
e^{ - \beta E_S(\Gamma^\prime)}} ={ e^{- \beta E_S(\Gamma)} \over
{\cal Z}_S} 
\label{eqn:pg}
\end{equation}

\noindent 
evaluated at any convenient but sufficiently low temperature.  For
sequences with a unique ground state, any temperature below the
folding transition temperature, at which the probability of occupancy
of the native state is $1/2$, would suffice. 

A physical and rigorous approach to the design problem on a structure,
$\Gamma$ would consist of exploring both the family of sequences and
the family of conformations to identify a sequence (or sequences) that
maximizes the low-temperature occupation probability (\ref{eqn:pg}).
A brute force application of this procedure is not feasible for chains
of 48 residues and hence it becames paramount to find good
approximations to (\ref{eqn:pg}) to ease the computational effort.

The HSF strategy for design \cite{6,7,8} was based on limiting the
search in sequence space to chains having exactly $n_H=24$ hydrophobic
amino acids each and isolating those with the lowest possible energy
in the target structure. The HSF choice of setting $n_H=24$ was based
on the expectation that the selected putative solutions would have
minimal degeneracy (since the composition is intermediate between the
two trivial cases already mentioned, $n_H=0$ and $n_H=48$).

\begin{figure}
\vskip 1.0cm
\centerline{\psfig{figure=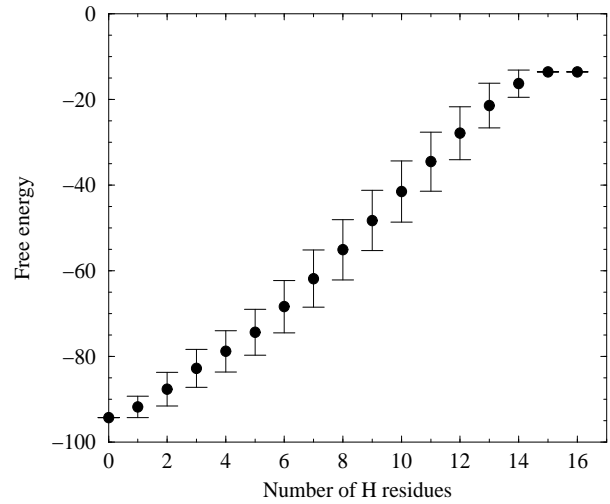,width=3.5in,height=3.0in}}
\caption{Plot of the average free energy as a function of the number 
of hydrophobic residues for two-dimensional chains of length 16 at 
$K_B T = 0.1$. The error bars denote the standard deviations.}
\label{fig:delta}
\end{figure}

A more quantitative insight into the validity of restricting $n_H$ to
equal the number of $P$ residues, $n_P$, can be obtained by studying
the composition of all good sequences (i.e., those with an unique
ground state) of length 16 in two dimensions, for which an exhaustive
search is feasible with modest numerical effort. It turns out that the
value of $n_H$ for the good sequences ranges from as little as 4 up to
14, while the number of sequences with $n_H=n_P=7$ is less than 20
$\%$ of the total number of good sequences.  A further exact study of
16-monomer chains shows that, to a very good approximation, ${\cal
Z}_S$ does not vary significantly on considering sequences with the
same fixed H/P composition. This is illustrated in Fig.
\ref{fig:delta} which shows that ${\cal Z}_S$ can, in a zeroth order
approximation, be considered as a function of the single variable
$n_H$.

Inspired by these two observations, we proceeded as follows.  First we
decided to partition all sequences of length 48 in bins according to
their value of $n_H$ ($10 \le n_H \le 24$). Values of $n_H$ less than
10 were not considered because, as  already mentioned, the
degeneracy of the associated native state is, presumably, gigantic.
The reason for setting 24 as an upper limit for $n_H$ was dictated by
the fact that the complexity and CPU requirements of the CHCC
algorithm used to test the correctness of our answers grows rapidly as
a function of $n_H$. This limitation, is therefore not to be regarded
as intrinsic to our design strategy which, in fact, does not depend on
it.

\noindent The binning procedure is aimed at dividing the sequences
into homogeneous groups within which the partition function, ${\cal
Z}_S$ does not fluctuate wildly. By assuming a constant ${\cal Z}_S$
for all sequences in a bin the maximization of the ``occupation
functional'' (\ref{eqn:pg}) (restricted to the same bin) only requires
the identification of the sequences with maximum number of contacts on
the target structure. We performed this by doing a simulated
annealing in sequence space where the elementary move is a swap of one
or more H/P pairs of residues in the chain and typically isolating the
five best resulting sequences. 

\noindent The above step can also be regarded as a multibin extension
for the original HSF strategy (for which we have now provided a
quantitative justification). However, while remaining within the
constant ${\cal Z}_S$ approximation, it is not possible to compare the
different occupation probabilities for sequences in the same bin, and
especially across bins.

In order to carry out this comparison, we then devised a Monte Carlo
procedure for the calculation of  ${\cal Z}_S$.
The routine is based on the
possibility of expressing ${\cal Z}_S$ in terms of average quantities,

\begin{equation}
{\cal Z}_S \equiv \sum_\Gamma {e^{- \beta E_\Gamma(S)}} =  
{C_{tot}   \over \langle {e^{\beta E_\Gamma(S)}} \rangle_\Gamma}
\label{eqn:av}
\end{equation}

\noindent where the brackets denote an average taken over all
conformations and $C_{tot}$ is the number of self-avoiding walks of
length 48. Since $C_{tot}$ does not depend on $S$, it plays no role and can be
set equal to 1. This has only the effect of scaling by
the same numerical factor the occupation probability of all sequences.

For each selected sequence the average in (\ref{eqn:av}) was typically
taken over 100,000 independent conformations. This set of structures
was generated using the importance sampling method described in
ref. 11.  The engine of this procedure is the dynamical
construction of conformations by laying down successive portions of
the chain (typically made of 6-7 residues) in energetically favourable
positions. The construction follows a set of stochastic rules which
ensure that the resulting conformation is generated with the
appropriate probability according to Boltzmann statistics. 

\noindent The fictitious thermal energy scale, $1/\beta$, in
(\ref{eqn:av}), was set to 0.1, measured in units of the coupling
strength, $\epsilon_{HH}=1$. We found that, around this temperature,
the efficiency of the importance sampling algorithm remains
acceptable, while the lowest energy states are sampled with a
significant weight.  Then, for each selected sequence, the occupation
score was calculated as in (\ref{eqn:pg}) and the sequence with
highest weight was chosen as the putative solution. The calculation of
${\cal Z}_S$ for the 75 selected sequences took, on average 20 hours
of CPU time on a DEC Alpha workstation (while the CPU time required
for the selection of the 75 sequences was only of the order of
minutes). The list of the putative solutions to each ot the 10 HSF
structures is given in Table II.

The correctness of our answers was confirmed with the aid of the CHCC
algorithm \cite{24,25}. This technnique is based on the systematic
construction of all possible compact hydrophobic cores for a given
sequence. Later, the analysis of the compatibility of the core
geometry with the detailed composition of the sequence is performed.
The CHCC method also yields, as a by-product, a lower bound on the
ground state degeneracy of the solutions. For the sequences of Table
II this lower bound turned to be of the order of $10^3-10^4$, similar
to the ground state degeneracy of the original HSF putative solutions
\cite{6}.  The non unique encoding of the HSF structures is a
well-known undesirable feature of the HP model. In this respect, the
discreteness of the lattice and the limited number of residue classes
is detrimental to the degeneracy of the solution. On the other hand,
these very same features, allow the possibility to perform a robust 
check of a design procedure. 

For each of the 10 HSF conformations, additional correct
sequences were identified by our design procedure but were assigned a
lower $P_\Gamma(S)$ than for the one listed in Table II. For example,
for sequence PHHPPPHHHHHPPPHHHPPPPPHHPPPHPHPPPHHHPPHHPPHHPPHH ($n_H =
23, n_c =30$), which admits structure n. 7 as one of its native
states, the ground-state occupation probability was slightly
lower, but yet comparabale  within  error bars, than sequence 7 (Table
II).

Finally, Table II shows that there are a few extreme cases where the
solutions have a very low value of $n_H$. We believe that this may be
ascribed to the non-optimal designability of some of the HSF target
structures, which, in fact, were chosen at random among the
self-avoiding walks filling a 4x4x3 parallelopiped.

To summarize, we have presented a novel strategy to perform the inverse
design on three-dimensional lattice structures within the HP
framework. The method involves two steps of increasing numerical
difficulty and refinement in order to weed out bad sequences.
The method was tested on the Harvard-San Francisco problem 
\cite{6} for which a correct answer was obtained in all 10 cases.

Acknowledgements: We are indebted to Ken Dill and Eugene Shacknovich for
useful comments on the manuscript.  
This work was supported in part by INFN sez. di
Trieste, NASA,  NATO and the Center for Academic Computing at Penn
State.

\vskip 3.0cm
\begin{table}
\begin{center}
\begin{tabular}{|r|c|}\hline
 & Structures  \\
\hline \hline
1  & {\tt RFFRBULBULDFFFRBUFLBBRRFRBBLDDRUFDFULFURDDLLLBB}  \\
2  & {\tt RFFLBUFRBRDBRFFFLBUULBLFFDDRUURDRUBDBULBRDLLULD}  \\
3  & {\tt RFUBUFFRBDFDRUUBBUFFLLLDDRDLBUBUFUBRFRBDDRFDBLF}  \\
4  & {\tt RRFLLUFDRUUFRBBBDLLURFDRFDFRBUBDBUUFFFDLLDLUUBB}  \\
5  & {\tt RFLFUBBRFFFLDRBRUULBRDDRBLUULLFFFRRRBBBDFFFLDRB}  \\
6  & {\tt RFFRBUUULFDBBDLUFFUBBRRRFFLDRBBLDDRUFDFULLBLFDB}  \\
7  & {\tt RRRFULDLUULFURDDLDBUBRULUFRBRRDDLUFRULFRDLDRDLL}  \\
8  & {\tt RRRFLUUFDRDFULLBLBBUFFRBDDLFRRFLLUURRRBBDBLLURR}  \\
9  & {\tt RRFFRUULDLLUUBRFDBRDBRDFUUBUFFLBBDLULDDRFDFLBUU}  \\
10 & {\tt RFLFUUUBRBLDFRRBRFFUBBLFFLDRDLDRBRFUBBDLUFLLBRU}  \\
\end{tabular}
\vskip 0.5cm
\caption{The 10 compact self-avoiding conformations on a cubic lattice
introduced by the HSF group \protect{\cite{6}}. The
conformations are encoded in bond directions: U, up; D, down; L, left;
R, right; F, forward; B, backward.}
\end{center}
\end{table}

\vskip 2.0cm
\begin{table}
\begin{center}
\vskip 0.5cm
\begin{tabular}{|r|c|r|r| } \hline
 & Sequences & $n_H$ & $n_c$ \\
\hline \hline
1 &  {\tt PPHHHPPHPPPPHHPHHHHPHPHPHPPPPPPPPPPPPHHHPPPHHPHH} & 20 & 25 \\ 
2 &  {\tt HHHPPHHPPHHHHPHPPPPPPPHPPPPPPPPPPPPPPHPHHPPHHHPH}  &  18 &    22 \\
3 &  {\tt PPPHPPHHHHHHPPHHHPPHHHHPHHHPPPHPPHPPPHHPPPPHPPPP}  &  22   & 28 \\
4 &  {\tt PPHHPPPPPPPPPPPHPHPPPPPPHPPPPPPHHHHPHPPPPPPPPPPP}  &  10 &    11 \\
5 &  {\tt PPPPPHPPPPHHPPPPPHHHPPPPPPPPPPPPHPHHPHPPPPHPHPPP}   &  12 & 14  \\
6 &  {\tt PPPPPPHHPPPHHPHPPPPPPPPPPPPPHPHPPHPPPHPPPHPHPPPP}  & 11  &   13   \\
7 &  {\tt PPPPPPHPPHHPPPHHHPPPPPHHPPPHPHPPPHHHPPHHPPHHPPPP}  & 17  &   22   \\
8 &  {\tt PHHPHHHPHHHHPPHHHPPPPPPHPHHPPHHPHPPPHHPHPHPHHPPP}  & 24  &   31   \\
9 &  {\tt PPPPHPPPPHHPHPPPPHHPHPPPPPPPPPPPPPPPPPPPPHPHPPPH}  & 10  &   11   \\
10&  {\tt PPPPPHHPPPPPPHHPHPPPPPPPPPPHPPHPPPPPPPPPPHPHHPHH}  & 12 & 14 \\
\end{tabular}
\vskip 0.5cm
\caption{The 10 HP sequences selected by our design procedures as
having the highest occupation probability, $P_\Gamma(S)$, in the
corresponding conformations of table I.  $n_H$ is the number of
H-amino acids in the sequence and $n_c$ is the number of H-H pairs
that are not next to each other along the chain but yet are nearest
neighbors in the target structure.  We have confirmed that the design
procedure is succesful in all the 10 cases by using a constraint-based
hydrophobic core construction (CHCC) algorithm.
[\protect{\onlinecite{24,25}}]}
\end{center}
\end{table}

\vspace{0.2cm}

\end{document}